\documentstyle[aps,preprint,epsf,array]{revtex}

\def\gsim{\mathrel{\raise.3ex\hbox{$>$\kern-.75em\lower1ex\hbox{$\sim$}}}}
\def\lsim{\mathrel{\raise.3ex\hbox{$<$\kern-.75em\lower1ex\hbox{$\sim$}}}}

\draft
\begin{document}

\vspace*{1cm}

\begin{center}
{\Large {\sc  Bulk Properties of Anharmonic Chains in Strong Thermal 
              Gradients: Non-Equilibrium $\phi^4$ Theory}}

\vspace{1.3cm}

Kenichiro AOKI$^a$\footnote{ E--mail: {\tt ken@phys-h.keio.ac.jp}}
 {\rm and} Dimitri KUSNEZOV$^b$\footnote{E--mail: 
{\tt dimitri@nst4.physics.yale.edu}} \\

 $^a${\small\sl Dept. of Physics, Keio University, 
    4---1---1 Hiyoshi, Kouhoku--ku, Yokohama 223--8521, Japan}\\
  $^b${\small\sl Center for Theoretical Physics, Sloane Physics Lab, Yale
  University, New Haven, CT\ 06520-8120}

\vskip 1.8 cm
\parbox{13.0cm}
{\begin{center}\large\sc ABSTRACT \end{center}
{\hspace*{0.3cm}
  We study nonequilibrium properties of  
  a one-dimensional lattice Hamiltonian with quartic interactions
  in strong thermal  gradients.  Nonequilibrium temperature
  profiles, $T(x)$, 
  are found to develop significant curvature 
  and boundary jumps. From the determination of the bulk thermal
  conductivity, we develop a quantitative description of $T(x)$
  including the jumps.}}
\end{center}

\vspace{5mm}

\noindent PACs numbers:   05.70.Ln, 05.60.-k, 44.10.+i, 02.70.Ns

\noindent keywords:  Non-equilibrium steady state, thermal conductivity,
                long-time tails, Green-Kubo, anharmonic chains,
                Fourier's law.

\newpage

The description of transport in physical systems is usually
relegated to the near equilibrium regime, where Green-Kubo
theory can be used. When one strays from this into systems far
from equilibrium, far less is known about the statistical
mechanics, or the behavior of transport
coefficients\cite{hoover}. 
While the physics of non-equilibrium systems is 
certainly of broad interest, many basic problems have yet to be fully
understood. Of particular interest are the non-equilibrium
stationary states and the nature of the statistical mechanics
which characterize it. One approach to understanding this
problems is to place systems in thermal gradients and examine
the long time behavior of observables. In one dimension, this
has been done in a variety of problems where both finite and
divergent transport coefficients were measured. Typically the
transport coefficients are found to diverge when the Hamiltonian
conserves total momentum. This is typical of systems where the
interactions depend only upon differences $x_i-x_j$,
such as the FPU and Toda
chains\cite{hoover,fpu-later,div-transport}.
When an `on site' potential, $V(x_i)$, is also present,
the coherent propagation of long wavelength modes is
suppressed resulting in finite conductivity\cite{hata}. This
is the case in the Frenkel-Kontorova,
ding-a-ling, Lorenz and other models\cite{finite-1d-transport,tprofs,casati,takesue}. 
Bulk behavior is also known in higher dimensions as well\cite{higher}.

In this letter, we investigate the
non-equilibrium steady state properties and thermal transport 
of lattice Hamiltonians with
quartic interactions in 1 spatial dimension. Our system
is the discrete version of $\phi^4$ scalar field theory, a
proto-typical field theory that has broad application.  In contrast to 
systems such as the FPU
$\beta$ model\cite{fpu-later,hata,ford}, which have divergent thermal 
conductivities,
we find a well defined bulk limit for our non-equilibrium
results. This we attribute to the on-site nature of the $\phi^4$
interaction. Certain properties of this theory have been studied
in the past, which include the Hamiltonian dynamics and phase
transitions, as well as the ergodic
properties\cite{phi4}. However, the thermal conductivity has not
been determined, and further, there is yet no understanding of
the role of boundary jumps and its inter-relation with the shape
of the non-equilibrium thermal profile.
 We present a quantitative analysis of this effect which
describes the behavior near and far from thermal equilibrium.
A full description of the temperature profiles far from
equilibrium is shown to require a description of the boundary jumps.

Our model system is the 1-dimensional Hamiltonian

\begin{equation}  \label{lattice-hamiltonian}
  H'=\frac{1}{2}\sum_{i=1}^L\left[\frac{\hat p_i^2}{m}  
          + \chi^2(\hat q_{i+1}-\hat q_i)^2 
          + \frac{1}{2}\beta^2 \hat  q_i^4\right],
\end{equation}
where $\beta^2$, $m$ and $\chi$ are parameters. This model becomes
identical to the FPU $\beta$ model if we substitute $(q_{i+1}-q_i)^4$
for $q_i^4$ in the interaction. It is convenient to
perform the rescalings $\hat q_i = q_i (\chi/\beta)$, $\hat p_i = p_i
(\sqrt{m}\chi^2/\beta)$ and $\hat t= t (\sqrt{m}/\chi)$. In the
dimensionless variables $q_i,p_i,t$, the Hamiltonian is

\begin{equation}  \label{lattice-h}
  H=\frac{1}{2}\sum_{i=1}^L\left[p_i^2 + (q_{i+1}-q_i)^2 +\frac{1}{2}
              q_i^4 \right],
\end{equation}
where $H= H' \beta^2/\chi^4$. To study the statistical mechanics of $H$ 
in strong thermal gradients, we evolve Hamilton's equations of
motion together with thermal boundary conditions on the two
ends. Observables are allowed to converge to steady-state values,
which are then analyzed. The dynamics is solved on a spatial grid 
using either fifth and sixth
order Runge-Kutta or leap-frog algorithms\cite{numrec}. We impose
time-reversal invariant, thermal boundary conditions on the equations
of motion at sites $i=1$ and $i=L$. Specifically, the endpoint equations are 
augmented to include dynamical thermal constraints using the robust methods of
\cite{thermostats,sloan}, and become:

\begin{equation}
\begin{array}{lcl}
\dot q_1 & = & p_1,\\
\dot p_1 & = & -\frac{\partial H}{\partial q_1} -
  \frac{a_1}{T_c^0}w_1^3 p_1 - 
  \frac{a_2}{T_c^0}w_2 p_1^3,\\
\dot{w}_1 & = &a_1( p_1^2/T^0_c-1),\\
\dot{w}_2 & = & a_2(p_1^4/T^0_c - 3  p_1^2)
\end{array}
\qquad 
\begin{array}{lcl}
\dot q_L &=& p_L,\\
\dot p_L &=& -\frac{\partial H}{\partial q_L} -
   \frac{a_3}{T_h^0}w_3^3 p_L 
 - \frac{a_4}{T_h^0}w_4 p_L^3,\\
\dot w_3 &=& a_3(p_L^2/T^0_h-1),\\
 \dot w_4 &= &a_4(p_L^4/T_h^0 - 3  p_L^2).
\end{array}
\end{equation}
The auxiliary variables $w_k$ dynamically implement the statistical
boundary conditions consistent with the boundary temperatures $T^0_c$
and $T^0_h$\cite{thermostats}. We will use the optimal choices
for the couplings $a_i$, which in this case are $a_i=1$\cite{sloan}.
( We have checked that the results presented here do not
depend on this particular choice of the
thermostats $w_k$, the values of $a_i$, nor on the number of sites
at each end that are thermostatted. We have also examined
thermostatting from 1 to 3 sites on each end; we return to this point when
we discuss boundary jumps.)
 With this control of the endpoint temperatures, we are able
to simulate the dynamics of $H$ with the boundary thermostat
temperatures $(T^0_c, T^0_h)$ fixed.  Apart from the endpoints, the
system evolves according to the dynamics dictated by the Hamiltonian
(\ref{lattice-h}). In the simulations, we use between $10^6$
and $10^9$ with time steps of $dt$ from $0.1$ to $0.001$. The lattice
size was varied from $L=10$ to 8000.

In order to understand the transport properties, it is useful to start 
with the stress-energy tensor ${\cal
T}^{\mu\nu}$. From the continuity equation,
\begin{equation}
 \frac{\partial}{\partial x^\mu}{\cal T}^{\mu\nu} 
 =\frac{\partial}{\partial t} E + \nabla  J=0,
\end{equation}
where $E={\cal T}^{00}$ is the energy
density. For thermal gradients in the $x-$direction, we can identify 
$J={\cal T}^{0x}(x_i,t)=-p_i(q_{i+1}-q_i)$ as the heat flux.

\noindent \underline{$T_c^0=T_h^0$ :} By setting $T_c^0=T_h^0=T$,
we verify that the canonical ensemble is realized at all
sites. This is done in several ways. By following the trajectories
$p_i(t)$ and histogramming the values at each time step, one
reconstructs the entire momentum distribution at any site, 
which are found to converge to
$f(p_i)\propto \exp[-p_i^2/2T]$. We also verify the virial theorem for
the kinetic and potential energies.  

In thermal equilibrium, we can use the Green-Kubo formula to compute
the thermal conductivity $\kappa(T)$ in the linear-response regime:

\begin{equation}  \label{green-kubo}
  \kappa(T)={\frac{1}{N T^2}}\int_0^\infty\!\!dt\sum_{k,k'}
  \left\langle J(x_k,t)
  J(x_{k'},0)\right\rangle_{\scriptscriptstyle EQ}.
\end{equation}
The autocorrelation function is evaluated in the canonical ensemble,
$T^0_1=T^0_2$, and the number of points used in the sum is $N<L$ since we omit
points near the boundaries.  The formula is expected to hold within
the linear regime, at least in higher dimensions.  In 1-d, the
integrand in (\ref{green-kubo}) has been argued to develop a long time
tail $\sim t^{-1/2}$ on time scales on the order of a few ten times
the mean free time, leading to the divergence of
(\ref{green-kubo})\cite{tail}.  In our case, the presence of the
on site interactions serve to destroy this long-time tail at
sufficiently large times.
We plot the absolute value of the 
integrand of (\ref{green-kubo}) for a 
temperature $T=1/10$ in  Fig. 1~(top). At this temperature the mean
free time roughly 50, so that long-time tails should be present on
times $\sim 10^3$.  The
$t^{-1/2}$ behavior is indicated by the dashed line. The time
integral is given in 1~(bottom), showing that the integral (\ref{green-kubo})
is finite.  These linear response predictions will be seen to
agree with the direct measurements we discuss
below.  We see that on the time scales of $t\sim 50$, there is some
agreement with the long-time tail predictions. However, on longer
times the divergent behavior is not present, and the linear response
results converge to a well defined limit.
In Fig. 2, we compile the Green-Kubo measurements
of $\kappa$, plotted as a function of $T$. In the context of the
FPU model, it has recently been shown that the long-time tails
of the autocorrelation function behave as  $t^{-3/5}$ leading to a
$t^{2/5}$ divergence, based on mode-coupling
theory\cite{lep}. While we do not have a divergence in our integrals, 
the behavior of the integrated autocorrelation function in the
transient regime is much closer to $t^{1/2}$  at all measured
temperatures.

\noindent\underline{$T_h^0\gsim T_c^0$ :} We can verify the linear
response calculations through a direct measurement of the heat
flow. The thermal conductivity $\kappa$ is defined using Fourier's law
$\langle J\rangle_{{ \scriptscriptstyle NE}}= -\kappa \partial_x
T(x)$, where $\langle J\rangle_{{ \scriptscriptstyle NE}}$ is the
averaged heat flux. The gradient is evaluated by taking $|T_h^0-T_c^0|$
sufficiently small so that the temperature profile is
linear. We then vary the temperature difference
$|T_h^0-T_c^0|$ around the same average temperature to verify that 
 $J$ is proportional to $\nabla T$, and extract $\kappa(T)$.
$T(x)$ is the local temperature defined through an
ideal gas thermometer, by $T(x)=\langle p^2(x)\rangle_{{
\scriptscriptstyle NE}}$, where $p(x)$ is the momentum  at site $x$.
Here $\langle\cdots\rangle_{{ \scriptscriptstyle NE}}$ indicates the
ensemble average over the non-equilibrium steady state. To obtain the
transport properties, each simulation is run long enough for
observables such as $J$, the energy density as well as
distribution functions to converge. 
In Fig. 2 we show the results of both direct measurements and
Green-Kubo
predictions for  $\kappa$. We find that both agree over several orders 
of magnitude, and that $\kappa(T)$ 
has the power law temperature dependence 
\begin{equation}\label{kappa}
  \kappa(T)=\frac{A}{T^{\gamma}},\qquad\qquad \gamma=1.35(2),\qquad
A=2.83(4),
\end{equation}
reminiscent of the
behavior of lattice phonons at high temperature\cite{phonon}. We
have also verified that a sensible bulk behavior exists, as
shown in Fig. 3; the thermal conductivity is
independent of $L$ when it is larger than the mean free path,
which, on the lattice, is of order of the conductivity (see below). The dashed
lines in Fig. 3 correspond to the values of Eq. (\ref{kappa}) at that 
temperature.

\noindent\underline{$T_h^0\gg T_c^0$ :}
By controlling $T_c^0$ and $T_h^0$ we can begin to explore the 
non-equilibrium steady state.
For $T_c^0\sim T_h^0$, we expect to be in the linear regime, with a
linear temperature profile, and transport given by Fourier's law. This 
has been readily verified. As the
difference between endpoint temperatures increases,  two
characteristics emerge: the temperature profile develops
curvature (Fig. 4(a) where $T_c^0/T_h^0=0.05$), and there are
substantial temperature jumps near the
boundaries (Fig. 4(c)).

When the temperature varies substantially in the system, one
cause for the non--linearity is the temperature dependence of
the thermal conductivity. If we assume that the temperature dependence 
of the thermal conductivity is the dominant source of the
non-linearity of the thermal profile, $T(x)$ can be obtained by
integrating Fourier's law as
\begin{equation} \label{t-profile}
  T(x) = T_c\left[1-\left(1-\left(\frac{T_h}{T_c}
      \right)^{1-\gamma} \right)
    {\frac{x}{L}}\right]^{{\frac{1}{1-\gamma}}},\qquad \gamma\neq1.
\end{equation}
The integration is simple since the energy flow $\langle J\rangle_{NE}$ is
independent of $x$ due to current conservation. This result is distinct
from temperature profiles discussed previously as in Ref. \cite{tprofs}.
In this
equation, the temperatures $T_{c,h}$ denote the temperatures at
the boundaries obtained by extrapolating the temperature
profile. These are in general different from the endpoint
temperatures $T_{c,h}^0$. (In Fig. 4(c),
$T_c^0=0.05$ and $T_c=0.096$.) The temperature profile is 
a function only of $x/L$ so that it has a smooth continuum limit.
The ratio of Eq. (\ref{t-profile}) to the measured local temperature is shown in
Fig. 4~(b). Aside from the endpoints, one can see that the agreement
is quite good. We have found that this formula works well in higher
dimensions as well ($d=2,3$)\cite{ak1}. 

It is clear that while the formula for $T(x)$ describes the
shape of our observed non-equilibrium profiles, it depends on
quantities which are not determined from our input parameters,
namely the extrapolated temperatures. Hence a complete
description of $T(x)$ will require the understanding of how the
boundary jump $T_c-T_c^0$ (and similarly for $T_h$) depends on
the parameters of the system. We will do this below.

From Eq. (\ref{t-profile}), we find the heat flux in the
non-equilibrium steady state to be
\begin{equation}\label{ne-j}
  \langle J\rangle_{NE}= \frac{A}{L(1-\gamma)}(T_c^{1-\gamma}-T_h^{1-\gamma}).
\end{equation}
We can expand this around the average temperature $T_{av}=(T_h+T_c)/2$, to find:
\begin{equation}
 \langle J\rangle_{NE} =  \langle J\rangle_{NE}^0
       \left\{ 1+\frac{\gamma(\gamma+1)}{24}
      \left(\frac{\Delta T}{T_{av}}\right)^2 + 
      O\left(\left[\frac{\Delta
     T}{T_{av}}\right]^4\right)\right\},\qquad 
   \langle J\rangle_{NE}^0 =-\kappa(T_{av})\frac{T_h-T_c}{L}
\end{equation}
Here $\langle J\rangle_{NE}^0$ is identified as the constant
(temperature) gradient limit  of the heat flux, which actually 
provides a good approximation to the heat flux even when there is
curvature in the temperature profile, providing $\Delta T/T_{av}\lsim
1$. For $\Delta T/T_{av}\sim 1$, the notion of local
equilibrium becomes questionable, and this description of the heat
flux starts to break down\cite{ak1}.

As we mentioned previously, a common feature to non-equilibrium steady
states is the appearance of boundary jumps or
slips\cite{hata,fpu-later,pk}. These are not artifacts of the
simulation, but are true physical effects\cite{pk,exp}.  For systems
in thermal gradients, the boundary temperature can be different from
the temperature of the system near the edges. Similar behavior can
also appear in fluids which are sheared\cite{bhatt}.  In this case
there is slippage between the system and the wall. Such effects are
well known sources of error in the experimental measurement of the
transport coefficients in fluids\cite{exp}.  In our lattice model,
elementary kinetic theory predicts a thermal conductivity of
$\kappa=C_V c_s l$ for the system, where $C_V$ is the heat capacity
per volume, $c_s$ the sound velocity and $l$ the mean free path. The
formula applies to the basic excitations, the ``phonons'', of the
system. In our model, $C_V,c_s\sim1$.  Therefore, the mean free path
is expected to be of order of the conductivity itself.  This allows us
to clarify the nature of the temperature jumps at the boundaries. The
jump, arising due to the finite mean free path, satisfies a relation

\begin{equation}
 T_i-T_i^0=\eta \left.\frac{\partial T}{\partial
n}\right|_{boundary}=
\pm \eta \left.\frac{\partial T}{\partial x}\right|_{boundary}
\end{equation}
when $l\ll L$ \cite{pk}.  Here $\partial T/\partial n$ is the
normal derivative at the surface, so that the $+$ ($-$) sign
corresponds to the lower (upper) edge of the chain.
We will consider the lower temperature end (the $+$ sign), but
identical results hold for the high temperature side if one
changes the sign accordingly.
$\eta$ is expected to behave
like the mean free path and hence the conductivity, roughly
speaking. If we let $\eta=\alpha\kappa$ for some constant $\alpha$, 
we can use Eq. (\ref{ne-j}) to find the approximate behavior
of the boundary jumps:

\begin{eqnarray}\label{t-jumps}
 T_c-T_c^0 &\simeq &\alpha \langle J\rangle_{NE} \\
          &\sim  &  (T_h^0-T_c^0)\frac{\alpha
 \kappa(T_{av})}{L} + \cdots.\nonumber
\end{eqnarray}
We have determined $\eta$ 
by fitting Eq. (\ref{t-profile}) to profiles $T(x)$ obtained from various
non-equilibrium steady states, using the power $\gamma$ obtained from Eq.
(\ref{kappa}).
The behavior of $\eta$ as a function of $T$ 
from the non--linear thermal profiles is plotted in
Fig. 5 (top), which can be fitted to $\eta=(6.1\pm 0.5)\times
T^{-1.5\pm 0.1}$, consistent with
the above argument. We can also use (\ref{t-jumps}) to study the boundary
jumps directly.
In Fig. 5 (bottom), we see that Eq. (\ref{t-jumps}) is readily 
verified in the data, which
includes thermal profiles both near and far from equilibrium. A simple
fit (dashes) provides the coefficient $\alpha=2.6(1)$, fully
consistent with $\eta(T)=\alpha\kappa(T)$. We have analyzed these
effects in higher dimensions as well and find similar
results\cite{ak1}.

We now have the following behavior: Thermal boundary conditions will
result in a given amount of heat flux, which in turn determines the
magnitude of the boundary jumps through Eq. (11). Once the boundary
jumps are defined, the thermal profile, Eq. (7), is determined from
the endpoint temperatures $T_i^0$. We have examined the dependence of
the results we present here on the types of thermal boundary
conditions we have used. This includes variations not only of the
strengths of the interactions, $a_i$, but also of the number of
thermostatted sites on each end. We have observed that the coefficient
$\alpha$ (and hence $\eta$) is not a physical property of the
Hamiltonian, but can depend on the thermostats. Different thermostats
can produce different boundary jumps for the same boundary
temperatures $T_{c,h}^0$.  Although the new jumps provide new
extrapolated temperatures $T_{c,h}$, the temperature profile is still
predicted by Eq. (7). Hence the formulas we have presented are still
valid in spite of the (slight) differences in boundary jumps.  These
differences do not lead to any modification of the transport
coefficients, and are only differences in the shape $T(x)$ due
to the variation of the boundary jumps which seem to reflect the
efficiency at which heat can be transfered to the system through the
boundary in the far from equilibrium regime.

We have constructed the non-equilibrium steady states of an anharmonic
chain with quartic interactions, by placing it in strong thermal
gradients, and obtained well defined bulk transport properties.  These
results are also valid for classical $\phi^4$ lattice field
theory. We determined the temperature dependence of the
thermal conductivity, and then derived the non-equilibrium thermal
profile $T(x)$, which agrees well with the observed behavior in the
non-equilibrium steady state. A simple expression for the
non-equilibrium heat flux is also found to agree well with direct
measurements. Using these results, as well as arguments from kinetic
theory, we are able to quantify the temperature jumps at the edges of
our system, which are endemic to both non-equilibrium simulations and
experiment. We hope to develop this overall picture of the
non-equilibrium steady state towards understanding the dynamics of
phase transitions under non-equilibrium conditions, as well as the
concepts of local equilibrium.

We acknowledge support through the Grant--in--Aid from
the Ministry of Education, Science, Sports and Culture and
grants at Keio University and DOE grant DE-FG02-91ER40608.

\newpage
\begin{figure}
\begin{center}
  \epsfxsize=9cm\epsfbox{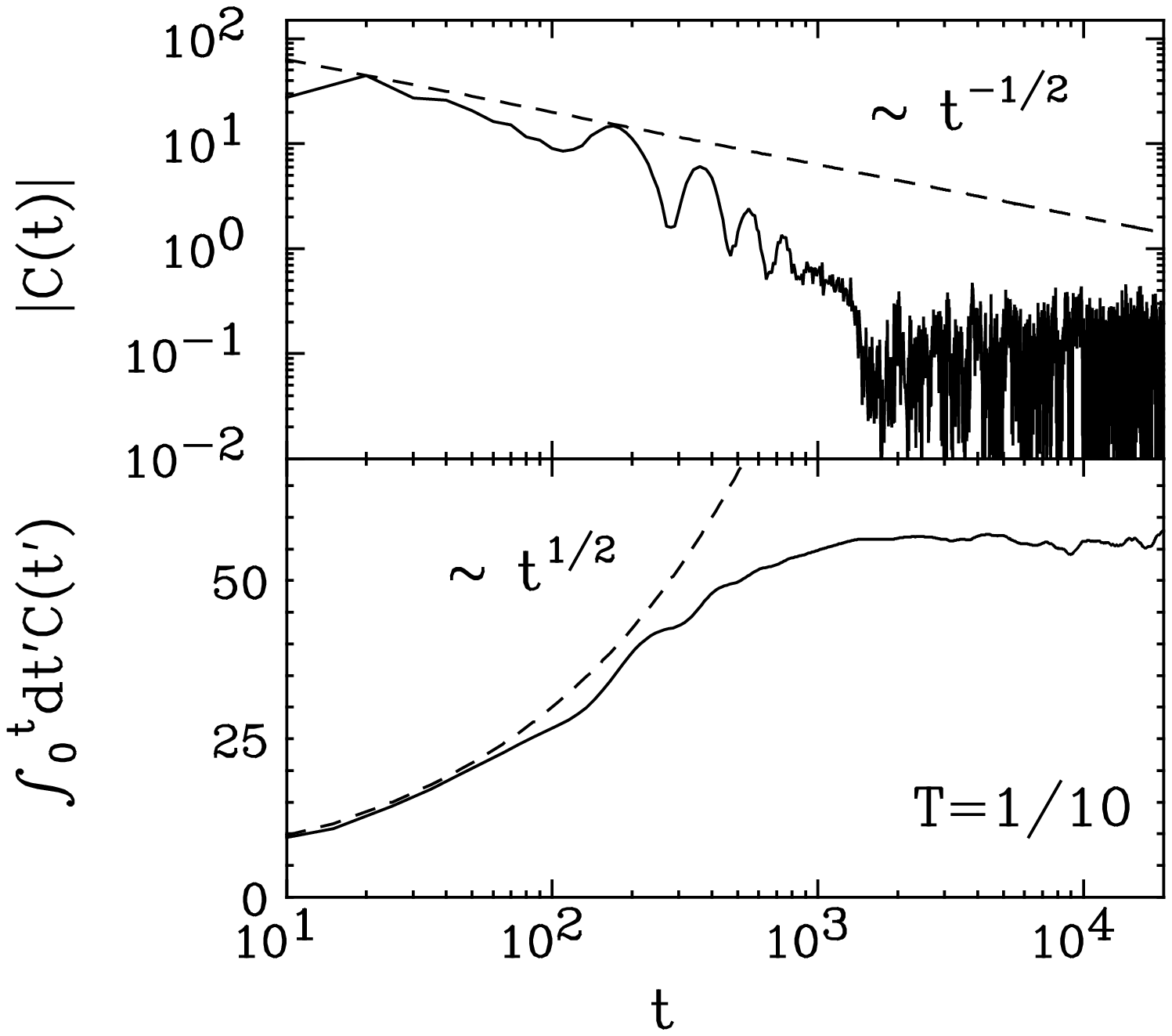}
  \caption{
    (Top) Absolute value of the autocorrelation function
    $C(t)=\sum_{k,k'} \langle J(x_k,t)
    J(x_{k'},0)\rangle_{\scriptscriptstyle EQ}/NT^2$, for $L=100$, up
    to a time $t=2\times 10^4$. The long-time tail divergence would
    have the behavior shown by the dashed line.  (Bottom) Green-Kubo
    integral up to time $t$, showing convergence of $\kappa$. The
    dashed line is the anticipated behavior if the long-time tail
    divergence were present. One can see that on the time scales
    $t\sim 10^3$ (a few ten times the mean free time), similar
    behavior can be seen, although it vanishes for larger times.}
\label{fig:thr}
\end{center}
\end{figure}

\begin{figure}
\begin{center}
  \epsfxsize=9cm\epsfbox{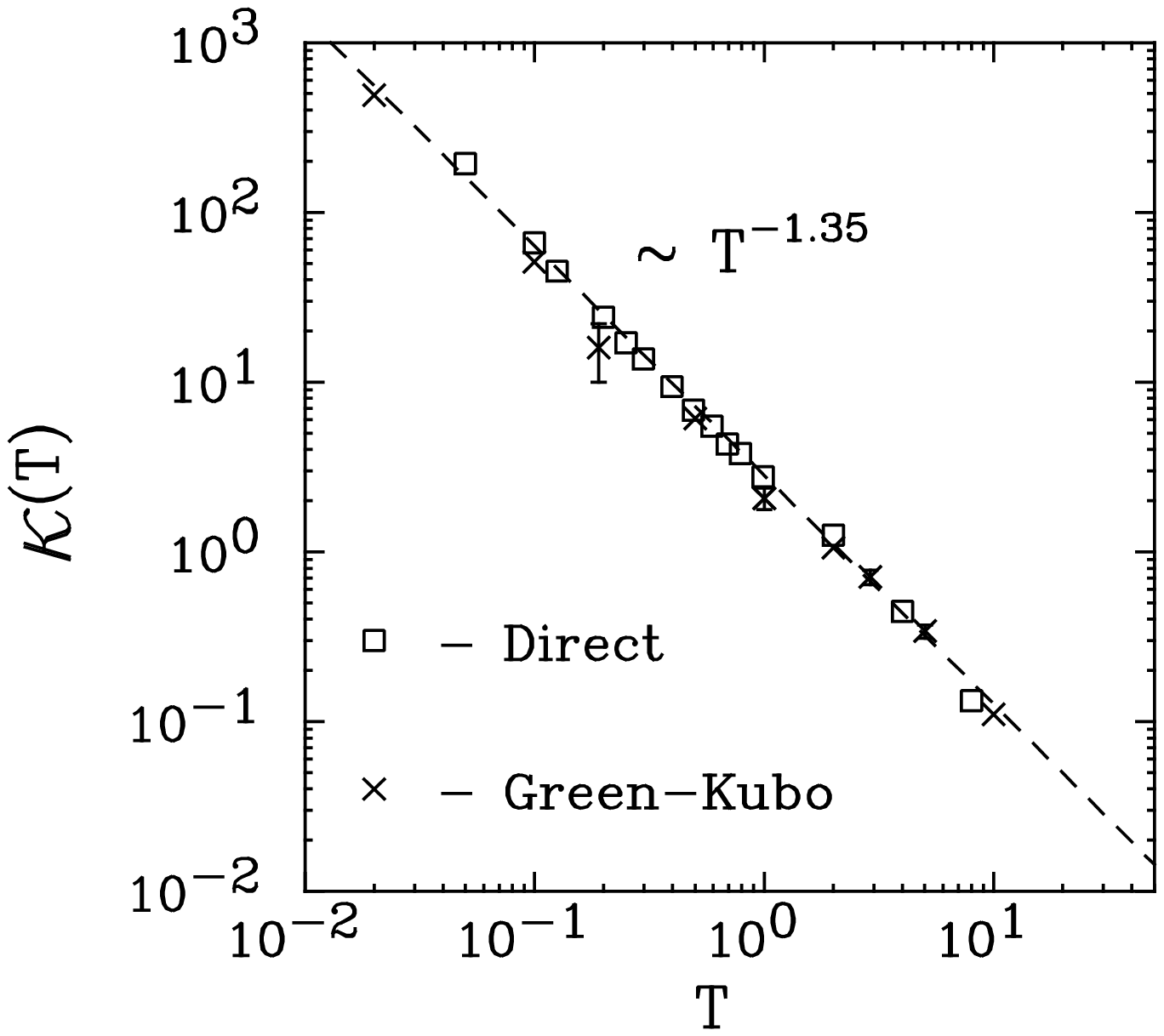}
\caption{ Thermal conductivity $\kappa$ obtained from
  direct ($\Box$) and Green-Kubo ($\times$) measurements for various
  lattice sizes $L$, and the power law fit
  $\kappa(T)=2.83(4)/T^{1.35(2)}$ (dashes). Green-Kubo integrals
  readily converge to values consistent with direct measurements. }
\end{center}
\end{figure}

\begin{figure}
\begin{center}
  \epsfxsize=9cm\epsfbox{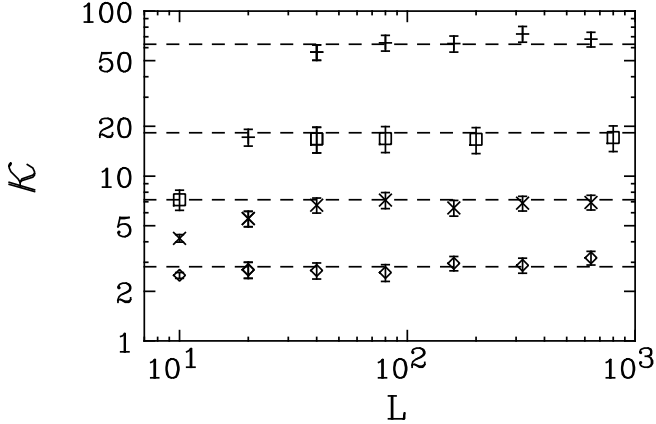}
  \caption{ Size ($L$) dependence of the thermal conductivity indicating
    bulk behavior, for temperatures (upper to lower) $T=1/10$ ($+$),
    $1/4$ ($\Box$), 1/2 ($\times$) and $1$ ($\Diamond$).  The dashed
    lines are the predictions from Eq. (6) for that temperature. }
\end{center}
\end{figure}

\begin{figure} 
\begin{center}
    \leavevmode
  \epsfysize=8cm\epsfbox{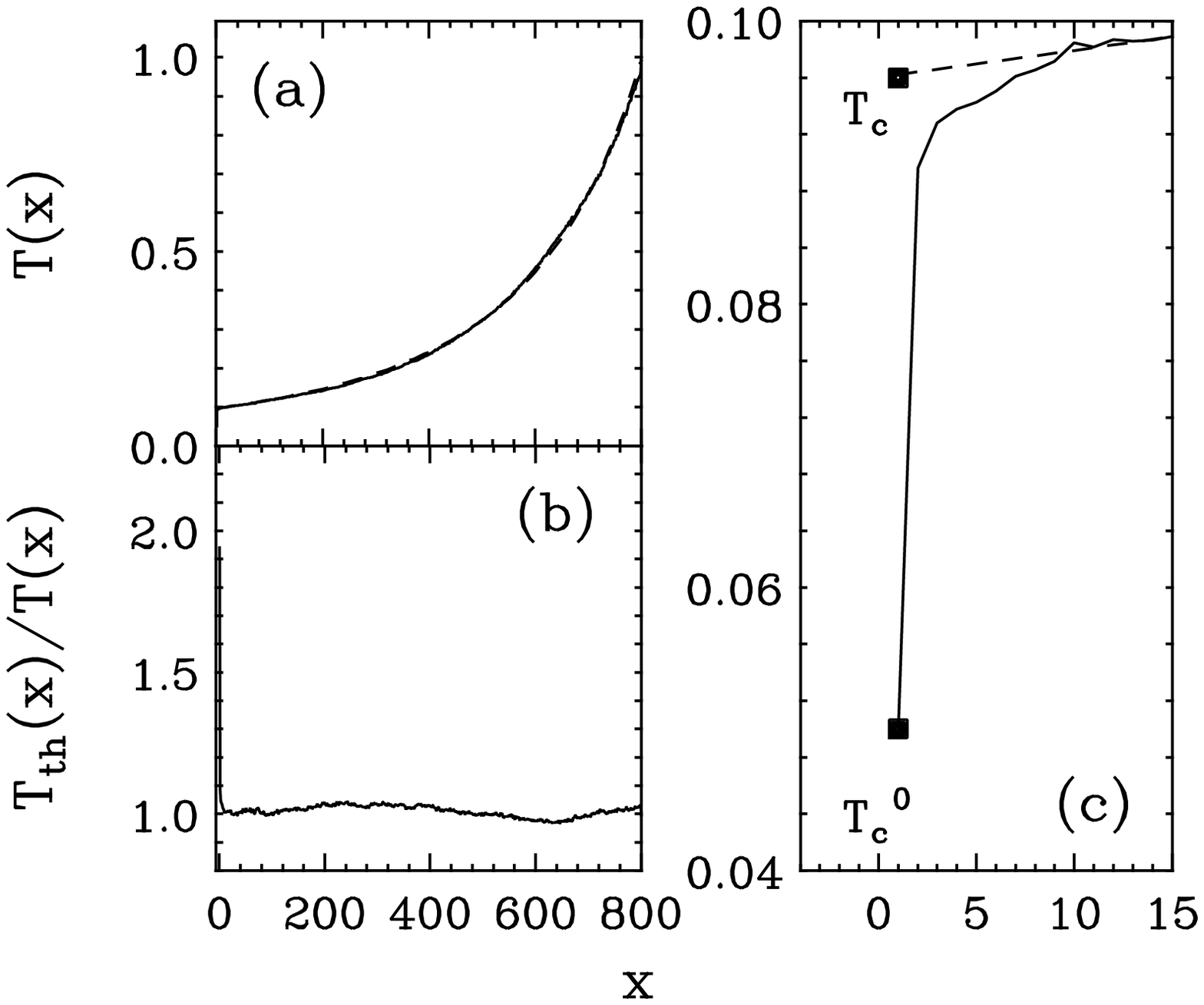}
  \caption{ (a) Non-equilibrium therm al profile $T(x)$ as a function of
    position (solid), compared to theoretical predictions of Eq. (7)
    (dashes) which lie on the solid curve. The temperature was sampled
    $10^6$ times every $\Delta t=0.25$.  Endpoints were thermalized at
    $(T_c^0,T_h^0)=(0.05,1)$. (b) The ratio of the predicted profile,
    Eq. (7), to that measured in (a). Aside from the boundary jumps,
    the agreement is within a few percent.  (c) Blowup of the
    temperature jumps at the low end of (a). The simulated temperature
    $T_c^0$ can differ significantly from the extrapolated value
    $T_c$.  }
  \end{center}
\end{figure}

\begin{figure}
\begin{center}
  \epsfxsize=9cm\epsfbox{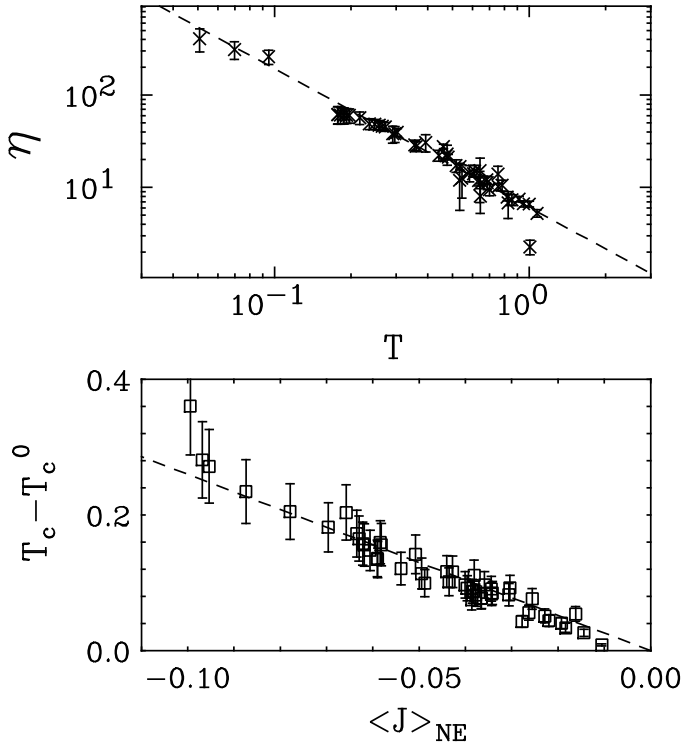}
  \caption{(Top) Temperature dependence of the jump parameter
    $\eta$ and its approximate power law behavior (dashes). (Bottom)
    Verification of Eq. (11). We plot the measured boundary jumps as a
    function of the non-equilibrium heat flux. The slope provides the
    coefficient $\alpha=2.6(1)$ which relates $\eta$ to the
    conductivity $\kappa$ via $\eta=\alpha\kappa$.}
\end{center}
\end{figure}
\end{document}